\documentclass{article}

\usepackage{arxiv}

\usepackage{graphicx}
\usepackage{amsmath}
\usepackage{amssymb}
\usepackage{amsfonts}
\usepackage{braket}
\usepackage[table,xcdraw]{xcolor}
\usepackage{subfig}
\usepackage{setspace}
\usepackage{multirow}
\usepackage[english, ruled, linesnumbered]{algorithm2e}
\usepackage{tikz}

\usepackage[T1]{fontenc}    
\usepackage{hyperref}       
\usepackage{url}            
\usepackage{booktabs}       
\usepackage{amsfonts}       
\usepackage{nicefrac}       
\usepackage{microtype}      
\usepackage{lipsum}		
\usepackage{natbib}
\usepackage{doi}

\newcommand\copyrighttext{%
\textcopyright 2021 IEEE. Personal use of this material is permitted. Permission from IEEE must be obtained for all other uses, in any current or future media, including reprinting/republishing this material for advertising or promotional purposes, creating new collective works, for resale or redistribution to servers or lists, or reuse of any copyrighted component of this work in other works. \url{https://doi.org/10.1109/BRACIS.2019.00149}.
}

\newcommand\copyrightnotice{%
\begin{tikzpicture}[remember picture, overlay, red]
\node[anchor=north, yshift=-20pt] at (current page.north) {\fbox{\parbox{\dimexpr\textwidth-\fboxsep-\fboxrule\relax}{\copyrighttext}}};
\end{tikzpicture}%
}

\title{Quantum Walk to Train a Classical Artificial Neural Network}

\author{ \href{https://orcid.org/0000-0002-6527-7065}{\includegraphics[scale=0.06]{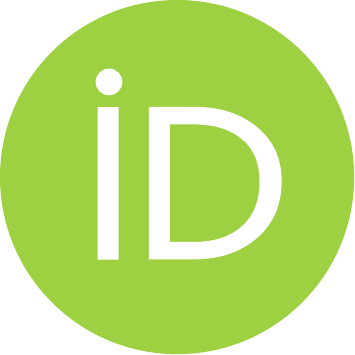}\hspace{1mm}Luciano S. de Souza}\thanks{95, R. Manuel de Medeiros, 35 - Dois Irmãos, Recife - PE.} \\
	Departamento de Estat\'{i}stica e Inform\'{a}tica\\
	Universidade Federal Rural de Pernambuco\\
	Recife, Brasil \\
	\texttt{luciano.serafim@ufrpe.br} \\
	\And
	\href{https://orcid.org/0000-0002-2672-7801}{\includegraphics[scale=0.06]{orcid.pdf}\hspace{1mm}Jonathan H. A. de Carvalho} \\
	Departamento de Estat\'{i}stica e Inform\'{a}tica\\
	Universidade Federal Rural de Pernambuco\\
	Recife, Brasil \\
	\texttt{jonathan.carvalho@ufrpe.br} \\
	\And
	\href{https://orcid.org/0000-0002-2131-9825}{\includegraphics[scale=0.06]{orcid.pdf}\hspace{1mm}Tiago A. E. Ferreira} \\
	Departamento de Estat\'{i}stica e Inform\'{a}tica\\
	Universidade Federal Rural de Pernambuco\\
	Recife, Brasil \\
	\texttt{tiago.espinola@ufrpe.br} \\
}

\date{September 1, 2021}

\renewcommand{\shorttitle}{Quantum Walk to Train a Classical Artificial Neural Network}

\hypersetup{
pdftitle={A template for the arxiv style},
pdfsubject={q-bio.NC, q-bio.QM},
pdfauthor={David S.~Hippocampus, Elias D.~Striatum},
pdfkeywords={First keyword, Second keyword, More},
}

\begin{document}
\maketitle

\copyrightnotice

\begin{abstract}
This work proposes a computational procedure that uses a quantum walk in a complete graph to train classical artificial neural networks. The idea is to apply the quantum walk to search the weight set values. However, it is necessary to simulate a quantum machine to execute the quantum walk. In this way, to minimize the computational cost, the methodology employed to train the neural network will adjust the synaptic weights of the output layer, not altering the weights of the hidden layer, inspired in the method of Extreme Learning Machine. The quantum walk algorithm as a search algorithm is quadratically faster than its classic analog. The quantum walk variance is $O(t)$ while the variance of its classic analog is  $O(\sqrt{t})$, where $t$ is the time or iteration. In addition to computational gain, another advantage of the proposed procedure is to be possible to know \textit{a priori} the number of iterations required to obtain the solutions, unlike the classical training algorithms based on gradient descendent. 
\end{abstract}

\section{Introduction}
\label{sec:introduction}

Quantum computing takes advantage of aspects of quantum mechanics to increase computational horizons \citep{yanofsky2008quantum}. Quantum effects, such as quantum parallelism, allow a computational gain and in many cases more efficient algorithms than its classic analog \citep{lloyd2013quantum}. For example, using quantum computing it is possible to search for elements in a disordered database in $\mathbf{}{O}(\sqrt{N})$ \citep{grover1996fast}.

The training of Classical Artificial Neural Networks (ANN) is performed fundamentally using the algorithm based on a descending gradient \citep{haykin}. It is possible to consider the problem of training a Neural Network as a search problem. The algorithm based on the descending gradient method searches for the minimum of an error function looking for an adequate configuration of the weights that allows the learning of the network \citep{biamonte2017quantum}.

Many efforts are currently being made in the search for efficient algorithms. In particular, there is a research field, named quantum machine learning \citep{wittek2014quantum}, where methods of machine learning and artificial intelligence are integrated on the quantum computing world in the hope to find more efficient procedures than the classical algorithms.

In particular, it is possible to find many works where a quantum artificial neural network or some quantum training procedure are proposed. For example, the work of \citet{zheng2018quantum} designed a method to train a perceptron using the Grover's search algorithm on a quantum computer~\citep{grover1996fast}. In \citet{ganjefar2017training}, a quantum neural network is proposed. This quantum neural network operates similarly to a classical neural network. \citet{patel2016novel}  created quantum binary neural network algorithm constructively. Efforts have also been employed in the development of procedures that use quantum algorithms in the training process of classical neural networks \citep{schuld2015simulating}.

Some works propose the use of quantum walking algorithm as search procedures \citep{wong2015faster,shenvi2003quantum,wong2015grover}. In this sense, we developed a computational procedure for the training process of classical artificial neural networks. Here, we are using the quantum walk algorithm in a complete graph, as proposed by \citet{wong2015grover}, as a search method for the synaptic weights to train the neural network.

This paper is organized as follows. In Section \ref{sec:quantum-computing} we introduce some concepts about quantum computation. In Section \ref{sec:quantum-walk}, we present the quantum walk. In Section \ref{sec:quantum-walk-in-complete-graph}, we present the quantum walk in complete graph used in this work for the development of the ANN training procedure. In Section \ref{sec:procedure}, we show the computational procedure proposed in this work. In Section \ref{sec:experiment}, we show the experiments performed. In Section \ref{sec:discussions} discusses the results obtained. Finally in Section \ref{sec:conclusion} is the conclusion.
\section{Quantum Computing}
\label{sec:quantum-computing}

Quantum computing is a computational paradigm, where there is a natural overlapping between computer science, mathematics, and physics branches. It takes advantage of some aspects of quantum mechanics to increase computational horizons~\citep{yanofsky2008quantum}. The basic unit of quantum information processing is called a quantum bit or qubit and describes a two-dimensional quantum system \citep{mcmahon2007quantum}. The possible states that a qubit can assume are represented by the state vectors presented in the Equation \ref{eq:qubit-representation}.

\begin{equation}
\label{eq:qubit-representation}
\ket{0} = \begin{bmatrix} 1 \\ 0 \end{bmatrix} \textrm{and} \ket{1} = \begin{bmatrix} 0 \\ 1 \end{bmatrix}\end{equation}

The qubit may be in one of two states $\ket{0}$ or $\ket{1}$, but may also exist in a superposition state characterized as a linear combination of these states $\ket{0}$ and $\ket{1}$. Considering $\ket{0}$ and $\ket{1}$ the orthonormal basis of a state space, an arbitrary state vector $\ket{\psi}$ can be defined as shown in Equation \ref{eq:estado-em-superposicao}.

\begin{equation}\label{eq:estado-em-superposicao}\ket{\psi} = \alpha \ket{0} + \beta \ket{1}\end{equation}

\noindent where $\alpha$ and $\beta \in \mathbb{C}$ are the probability amplitudes associated with states $\ket{0}$ and $\ket{1}$ respectively.

Although a qubit may be in a superposition state, whenever a measurement is performed, the state of the quantum system $\ket{\psi}$ will assume state $\ket{0}$ with probability $\left | \alpha \right | ^2 $, or state $\ket{1}$ with probability $\left | \beta \right | ^2 $. Since the squared module of these coefficients is related to the probability of obtaining a given result at the time of measurement, the following condition presented by Equation \ref{eq:soma-modulo-quadrado} must be respected.

\begin{equation}\label{eq:soma-modulo-quadrado}\left | \alpha \right |^2 + \left | \beta \right |^2 = 1.\end{equation}

In this way, the state of a quantum system is a vector $\ket{\psi}$ in a Hilbert space. The state $\ket{\psi}$ contains all the information we can get about the quantum system \citep{mcmahon2007quantum}. The evolution of this quantum system is described by a unitary transformation that depends on the application of an operator $U$ \citep{nielsen2002quantum}. To be unitary, a quantum operator $U$ must satisfy the condition shown in Equation \ref{eq:unitary-operator}.

\begin{equation}\label{eq:unitary-operator}
    UU^{\dagger} = U^{\dagger}U = I
\end{equation}
where $U^{\dagger}$ is your adjoint. This condition is necessary for the vector norm to be maintained, for this we must calculate $\parallel u \parallel \textrm{ = } \sqrt{\braket{u\mid u}}$, where $\braket{u\mid u}$ is the inner product.

The manipulation of quantum information is performed by operators, which are represented by unit matrices and act on the qubits, causing the temporal evolution of a quantum system \citep{yanofsky2008quantum}. Among the various quantum operators used to perform operations on qubits, there is the Hadamard port \citep{barbosa} represented by Equation \ref{eq:hadamard}.

\begin{equation}\label{eq:hadamard}H = \frac{1}{\sqrt{2}} \begin{bmatrix} 1 & 1 \\ 1 & -1 \end{bmatrix}\end{equation}

The Hadamard operator is one of the most important in quantum computing. Its application to a vector of the computational basis generates a state in superposition. This superposition phenomenon occurs when a quantum system state is into two or more of different pure states at the same time.
\section{Quantum Walk}
\label{sec:quantum-walk}

The quantum properties allow generalizing the concept of classical random walk \citep{marquezino2010analise}. Even retaining the idea of describing movements conditioned to random variables, the quantum walk generates a faster spreading when compared to classical one \citep{ambainis-one-dimensional}. 
In terms of computational complexity, the spreading of the quantum walk increases linearly with time $t$, that is, its variance is $ O(t)$, unlike its classic analog whose variance is $O(\sqrt{t})$.

In the one-dimensional case, the classical random walk can be understood as a particle which, depending on the outcome of a coin toss, will take a step to the left or to the right. But in the quantum approach, because of the superposition phenomenon, the particle can simultaneously take a step to the left and a step to the right.

According to \citet{portugal2013quantum}, in a generic way the quantum walk evolves under a space of Hilbert $\mathcal{H}_{M} \otimes \mathcal{H}_{P}$, in which $\mathcal{H}_{M}$ is the space of the coin that controls the movement of the walker and $\mathcal{H}_{P}$ defines the search space. For the one-dimensional case in discrete time, the coin space $\mathcal{H}_{M}$ is generated by the base $\{\ket{0}, \ket{1}\}$. The search space $\mathcal{H}_{P}$ is generated by the base $\{\ket{n} : n \in \mathbb{Z}\}$, which represents all the integers of the one-dimensional space.

The procedure evolves with successive applications of operator $U = S(H \otimes I)$, where $I$ is the identity, $\otimes$ is the tensor product and the shift operator $S$ is defined in Equation \ref{eq:shift-operator}.

\begin{align}\label{eq:shift-operator}
    S\ket{0}\ket{n} = \ket{0}\ket{n + 1}\textrm{, } S\ket{1}\ket{n} = \ket{1}\ket{n - 1}
\end{align}

\begin{figure}
    \centering
    \includegraphics[width=8cm]{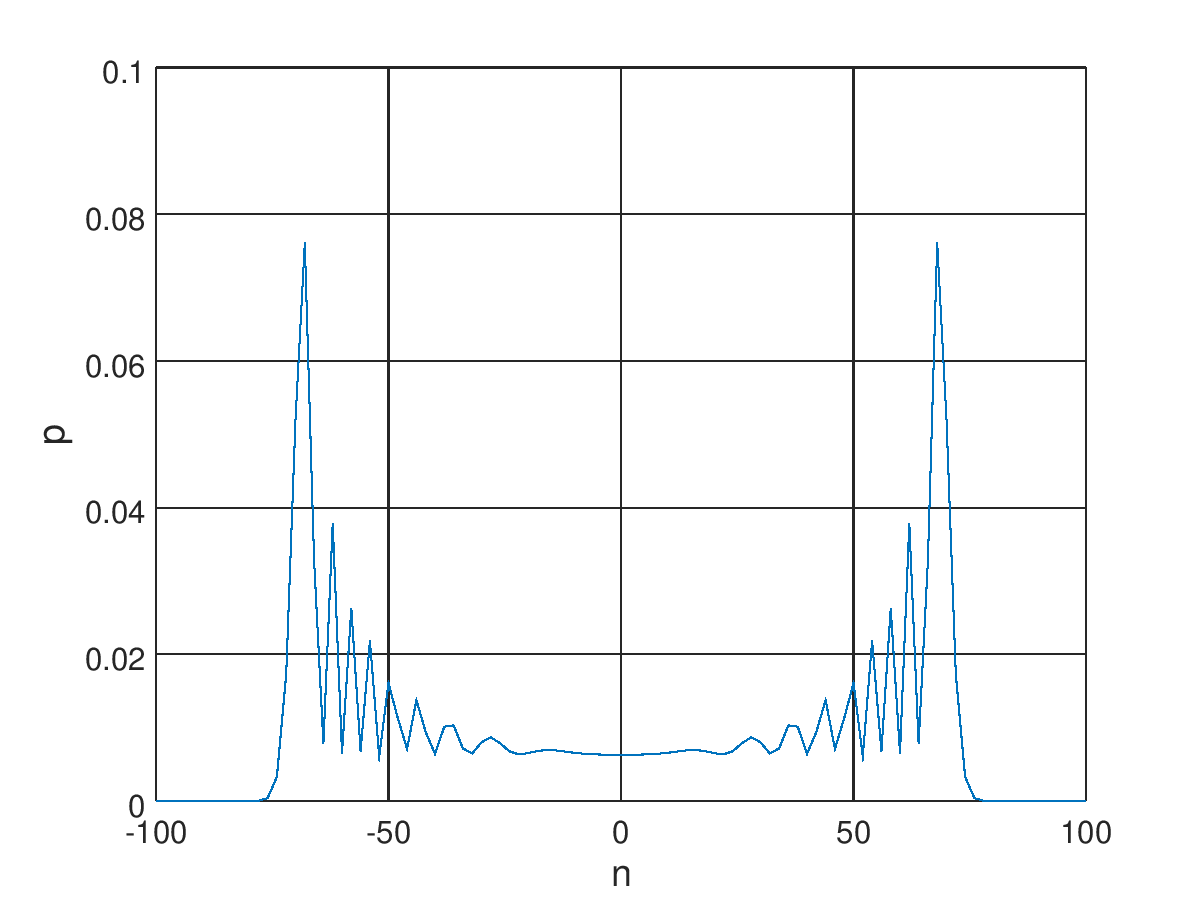}
    \caption{Probability distribution of the one-dimensional quantum walk after 100 steps.}
    \label{fig:distribuicao-de-probabilidade-caminhada-unidimensional}
\end{figure}

Therefore, for a given time $t$, the quantum walk state is given by $\ket{\psi(t)} = U^t \ket{\psi(0)}$. In Figure \ref{fig:distribuicao-de-probabilidade-caminhada-unidimensional} is shown the probability distribution at the end of one hundred applications of the operator $U$ ($t=100$), where the initial quantum state ($\ket{\psi(0)}$) is described in Equation \ref{eq:condicao-inicial-simetrica}. Unlike the classical case in which the distribution is a Gaussian centered on the origin, this quantum walk has a greater spread covering the interval $[-t/\sqrt{2},t/\sqrt{2}]$.

\begin{equation}
\label{eq:condicao-inicial-simetrica}
\ket{\psi(0)} = \frac{\ket{0}-i\ket{1}}{\sqrt{2}}\ket{n=0}
\end{equation}

In addition to the one-dimensional case, it is also possible to carry out the walk in a space of two dimensions. In this case, the computational basis for the coin space $\mathcal{H}_{M}$ is $\{\ket{i_{x},i_{y}} :  i_{x},i_{y} \in \{0,1\}^{2}\}$, whereas the search space $\mathcal{H}_{P}$ is now generated by $\{\ket{x,y} : x,y \in \mathbb{Z}\}$. The shift operator $S$ also needs to be changed to contemplate the new possibilities of movement of the walker, according to Equation \ref{eq:operador-de-evolucao-bidimensional}.

\begin{equation}
    \label{eq:operador-de-evolucao-bidimensional}
    S\ket{i_{x},i_{y}}\ket{x,y} = \ket{i_{x},i_{y}}\ket{x + (-1)^{i_{x}},y + (-1)^{i_{y}}}
\end{equation}

\begin{figure}
    \centering
    \includegraphics[width=8cm]{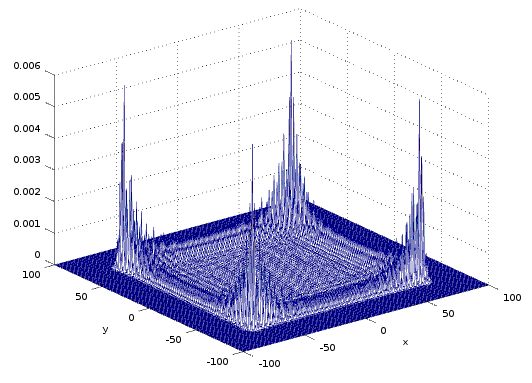}
    \caption{Probability distribution of the two-dimensional quantum walk after 100 steps.}
    \label{fig:distribuicao-probabilidade-caminhada-bidimensional}
\end{figure}

For this two-dimensional case, the probability distribution after one hundred iterations is shown in Figure \ref{fig:distribuicao-probabilidade-caminhada-bidimensional}.

\begin{figure}
    \centering
    \includegraphics[height=8cm]{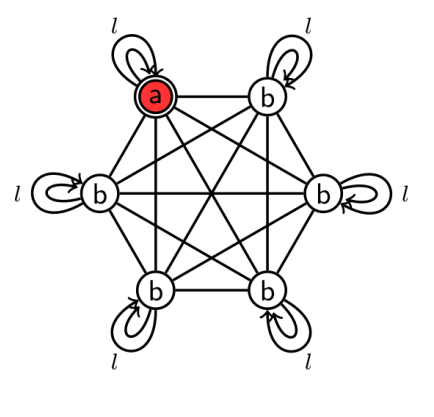}
    \caption{Complete graph with $N=6$ vertices. The single vertex marked as $a$ is indicated by the double circle and the red fill. Wong, 2015. Grover search with lackadaisical quantum walks.}
    \label{fig:grafo-completo-com-selfloops}
\end{figure}
\section{Lackadaisical Quantum Walk on Complete Graph}
\label{sec:quantum-walk-in-complete-graph}

Beyond the one-dimensional and two-dimensional quantum walks described in Section \ref{sec:quantum-walk}, variations of the technique can also be applied in other search spaces, such as the complete graph presented in work developed by \citet{wong2015grover}. In our graph representation, the vertices represent search space positions that will be converted into weights for the classical neural network. Because the graph is fully connected, the walker can move to any other location in a single step.

As can be seen in Figure \ref{fig:grafo-completo-com-selfloops}, the vertices are categorized into two distinct types. Those marked as $a$ solve the search problem, while those labeled as $b$ do not solve. For each of these categories of vertices there are two possible movements: move to another vertex of the same group, thus defining the quantum states $\ket{aa}$ and $\ket{bb}$, or move to a vertex of another category, represented by states $\ket{ab}$ and $\ket{ba}$. With the use of self-loops $l$ it is also possible to remain at the same vertex, so is also considered as a movement between vertices of the same category.

These quantum states are all formed by equal superpositions, as defined in Equation \ref{eq:definicao-base-quatro-estados}, where the vertices are represented by the states in the outer summation, while their corresponding edges are represented by the states in the inner summation of each relation. All this set of relations is valid only for the case of $l = 1$ self-loop at each vertex, $N$ is the total number of vertices of the graph and $k$ is the number of solutions.

\begin{align}
\begin{split}
\label{eq:definicao-base-quatro-estados}
\ket{aa} &= \frac{1}{\sqrt{k}} \sum_{a}^{}\left (\ket{a} \otimes \frac{1}{\sqrt{k}}\sum_{a'}^{}\ket{a\rightarrow a'}\right )\\
\ket{ab} &= \frac{1}{\sqrt{k}}  \sum_{a}^{}\left (\ket{a}\otimes\frac{1}{\sqrt{N-k}}\sum_{b}^{}\ket{a\rightarrow b}\right )\\
\ket{ba} &= \frac{1}{\sqrt{N-k}}  \sum_{b}^{}\left (\ket{b}\otimes \frac{1}{\sqrt{k}}\sum_{a}^{}\ket{b\rightarrow a}\right )\\
\ket{bb} &= \frac{1}{\sqrt{N-k}}  \sum_{b}^{}\left (\ket{b}\otimes\frac{1}{\sqrt{N-k}}\sum_{b'}^{}\ket{b\rightarrow b'}\right )
\end{split}
\end{align}

This more abstract representation is configured as a base change, causing the quantum walk to occur only in the space $\mathbb{C}^4$ and no longer in the complete initial graph. In Equation \ref{eq:estado-inicial-grafo-completo-redefinido} the initial state of the system is defined, which also evolves from successive applications of a unitary operator.

\begin{equation}
    \label{eq:estado-inicial-grafo-completo-redefinido}
    \ket{\psi_{0}} = \frac{1}{N} (k \ket{aa} + \sqrt{k(N-k)} \ket{ab} + \sqrt{k(N-k)} \ket{ba} + (N-k) \ket{bb})
\end{equation}

The maximum value of the probability of success is reached after the number of steps $t$ defined in Equation \ref{eq:qtd-passos-probabilidade-maxima}. It is worth mentioning that success is defined as the measurement of state $\ket{aa}$ or state $\ket{ab}$, since both represent the set of vertices marked as a solution.

\begin{align}
    \begin{split}
    \label{eq:qtd-passos-probabilidade-maxima}
    t = \frac{\pi}{\sqrt{2(2k+l-1)}}\sqrt{N}
    \end{split}
\end{align}
\section{Procedure for the training of neural network with lackadaisical quantum walk}
\label{sec:procedure}

The quantum walks are the basis for the method proposed in this work due to its capabilities as search algorithms. With the quantum walks, it is possible to perform an efficient search for the synaptic weights that train an Artificial Neural Network.

However, it can be seen in Figure \ref{fig:distribuicao-de-probabilidade-caminhada-unidimensional} that the amplitude amplification follows the spreading of the probability density function (PDF) over time. Thus, it is necessary to know the position of the desired state previously to calculate the number of steps for the quantum walk. With this number of steps, the natural quantum walk spreading will amplify the amplitude of that state, defining the exact moment of measuring the system.  As this information is not known \textit{a priori}, it is necessary to apply an approach capable of identifying along the process what are the desired states, and hence amplifying their respective amplitudes of probability.

Besides, the elaborated approach must allow a direct generalization of the search process, so that it can also be applied in situations with a higher amount of synaptic weights to be searched. Given all these needs, the quantum walk in a complete graph seemed adequate.

The vertices that solve the search problem are indistinguishable at the end of the process since they are all marked equally as $a$ in the graph. However, to generate the synaptic weights, it is necessary to obtain the specific location that the measured vertex represents. Because of this, the original procedure of the lackadaisical quantum walk was changed, where  the labels $\ket{\vec{r_i}}$ were included in the definition of the base states. These labels indicate the specific position that each vertex represents in the n-dimensional mesh. The new description of the base states is presented in Equation \ref{eq:redefinicao-base-quatro-estados}.

\begin{align}
\hspace{-0.195cm}
\begin{split}
\label{eq:redefinicao-base-quatro-estados}
    \ket{aa} &= \frac{1}{\sqrt{k}} \sum_{a}^{} \left( \ket{\vec{r}_{a}} \ket{a} \otimes \frac{1}{\sqrt{k}} \sum_{a'}^{} \ket{a\rightarrow a'} \right)
    \\
    \ket{ab} &= \frac{1}{\sqrt{k}} \sum_{a}^{} \left( \ket{\vec{r}_{a}} \ket{a} \otimes \frac{1}{\sqrt{N-k}} \sum_{b}^{} \ket{a \rightarrow b} \right)
    \\
    \ket{ba} &= \frac{1}{\sqrt{N-k}} \sum_{b}^{} \left( \ket{\vec{r}_{b}} \ket{b} \otimes \frac{1}{\sqrt{k}} \sum_{a}^{} \ket{b \rightarrow a} \right)
    \\
    \ket{bb} &= \frac{1}{\sqrt{N-k}} \sum_{b}^{} \left( \ket{\vec{r}_{b}} \ket{b} \otimes \frac{1}{\sqrt{N-k}} \sum_{b'}^{} \ket{b\rightarrow b'} \right)
\end{split}
\end{align}

Once this modification was made, it was possible to conceive the procedure of this work. The proposal is presented in the Algorithm \ref{alg:algoritmo-em-grafo-completo}. This algorithm covers the steps from the initial information collection and the preparation of the initial state, followed by the application of the quantum walk. After that, the state measurement at the end of the process is done, and the ANN's weights are adjusted.

\vspace{0.5cm}
\begin{algorithm}
\SetAlgoLined
\Begin{
Quantum count execution\\
Carry out base change\\
Preparation of initial state\\
\For{$j \leftarrow 1$ \KwTo $\frac{\pi}{\sqrt{2(2k+l-1)}}\sqrt{N} $}{
$\ket{\psi_{j}} \leftarrow U\ket{\psi_{(j - 1)}}$\\
}
Make a measurement in the states of the computational base $\{\ket{aa}, \ket{ab}, \ket{ba}, \ket{bb}\}$ \\
Carry out base change \\
Make a measurement in the states of the new computational base \\
Initialize the weights of the Artificial Neural Network
}
\caption{\textsc{Training Algorithm.}}
\label{alg:algoritmo-em-grafo-completo}
\end{algorithm}
\vspace{0.5cm}

In practice, the available computational resources define limitations on the size of the search space to be explored. Because of this, quantum counting~\citep{nielsen2002quantum} is performed on parts of the weight representation space, called windows, until one is found that contains at least one solution to the problem. This information of the exact number of solutions is required for the establishment of the new computational basis, as well as for the initial state and for the number of steps necessary to reach the maximum probability of success.

Then, the change to the new computational basis is made. In this step, an oracle must be used to determine which states are or are not a solution to the search problem. For the procedure proposed in this work, it is assumed that this oracle already exists, so just use it. With this new defined base, it is possible to prepare the initial state and perform the successive applications of the unit operator, according to the loop of lines 5 through 7. The speed-up claimed in this paper considers only the application of the quantum walk, so the complexity of Algorithm \ref{alg:algoritmo-em-grafo-completo} as a whole was not evaluated.

At the end of the quantum walk, according to line 8, a measurement is made in the states of the computational base $\{\ket{aa}, \ket{ab}, \ket{ba}, \ket{bb}\}$. With a high probability, one of the states containing the solution will be obtained, which can be the state $\ket{aa}$ or the state $\ket{ab}$. To return to the representation in the complete graph is necessary to perform the inverse of the base change made before the start of the walk, according to line 9.

As already mentioned, the goal is to obtain the label $\ket{\vec{r_i}}$ from one of the vertices that solve the problem. Following line 10, a new measurement must be performed, since the state obtained before the base change, $\ket{aa}$ or $\ket{ab}$, is a superposition of the vertices that are a solution. Finally, the weights of the neural network are adjusted with the result found in this last measurement, according to line 11.
\section{Experiment}
\label{sec:experiment}

An experiment was performed to verify the ability of the proposed procedure to train an artificial neural network, where the XOR classification problem was employed to it. In the following subsections, the main aspects that characterize the experiment are presented.

\subsection{Exclusive-OR Function}
\label{sec:exclusive-or-function}

The Exclusive-OR (XOR) function is a simple relation that maps two input bits to a single output bit, in such a way that, the output is $0$ when the input bits are equal while the output is $1$ when the input bits are different. As a classification problem, it is verified that no hyperplane can completely separate its classes, making the XOR function considered as a non-linearly separable problem (NLSP). Nevertheless, if a procedure is able to solve this simple problem, it means that the same procedure is also capable of solving more complex NLSP problems, which is the main reason why the XOR function was the target of this work.

\subsection{Neural Network Architecture}
\label{sec:neural-network-architecture}

The XOR problem can be solved using a Multilayer Perceptron (MLP) artificial neural network with two inputs, two neurons in the hidden layer and a single neuron in the output layer generating a single output. The logistic sigmoid was employed as the activation function~\citep{haykin} for all neurons. In total, the neural network has nine synaptic weights, six of them in the hidden layer and three in the output layer, already considering the bias of each neuron.

To train this neural network it is necessary to apply a procedure capable of finding nine values of synaptic weights. However, as this work investigated an abstracted two-dimensional quantum walk for a complete graph, at this stage it is not possible to find all the required weights.

Thus, it was applied a procedure inspired on extreme learning machine (ELM) \citep{tang2016extreme}. The seven weights between the initial and hidden layers were chosen randomly, remaining exactly two weights to be searched by the procedure proposed. It is worth note that this approach does not compromise the generality of the proposed methodology, where here the proof of conceit to apply a quantum walk to train an ANN was given. The quantum walk can be generated in any dimensionality, but the computational cost to execute a high dimensional quantum walk simulation grows exponentially with the number of dimensions.

\subsection{Weight Generation}
\label{sec:weight-generation}

In this work, the search is performed in quantum states that represent points in the space of integers. Therefore, the result obtained at the end of the process is formed by two integer values, one for each dimension of the space abstracted in the graph. But, the search space for the synaptic weights consists of the real numbers.

To perform this conversion of integers to real numbers, a real value $\Delta p$ was multiplied by each of the values obtained by the quantum walk. Then, assuming that the integers $x$ and $y$ were obtained after the measurement, the values of weights that would be defined for the neural network weights are $\Delta p*x$ and $\Delta p*y$, respectively.

In this way, the search is performed only between multiple values of that $\Delta p$. However, the value set for $\Delta p$ defines the refinement level of the search, behaving like the learning rate $\eta$ of the backpropagation algorithm.

\subsection{Measurement}
\label{sec:measurement}

The two measurement steps present in Algorithm \ref{alg:algoritmo-em-grafo-completo} were simulated using the roulette wheel selection \citep{taef2008}. This method is widely used in Genetic Algorithms, in which individuals of the population are chosen for the crossing phase based on their respective fitness values.

With a similar idea, the individuals in this work are the quantum states, while the fitness values are the probabilities of the system collapsing for each of the states. Experiments were performed to confirm the ability of this method to select quantum states according to their probabilities properly.
\section{Discussions}
\label{sec:discussions}

The computational procedure proposed in this work and presented in Section \ref{sec:procedure} has the purpose of searching for two synaptic weights that initialize the neural network together with the remaining set of seven fixed weights. In the Table \ref{tab:numeros-de-solucoes-por-janela-e-valor-de-deltap} is presented the quantitative solutions found by the procedure. For each $\Delta p$ value and window size, a different amount of solutions is obtained. When this value of $\Delta p$ decreases the search becomes more refined, increasing the chance of finding values that generate valid weights.

\begin{table}
\renewcommand{\arraystretch}{1.3}
\setlength{\tabcolsep}{15pt}
\centering
\caption{Number of solutions by window and $\Delta p$ values.}
\label{tab:numeros-de-solucoes-por-janela-e-valor-de-deltap}
\begin{tabular}{lccc}
\hline
\rule{0mm}{3mm} $\Delta p$ & 512x512 & 1024x1024 & 2048x2048 \\ \hline
0.05 & 52 & 52 & 52 \\
0.005 & 5202 & 5202 & 5202 \\
0.0005 & 19591 & 118649 & 485095 \\ \hline
\end{tabular}
\end{table}

To verify if the weights found by the procedure really train the neural network, these values were initialized in the network along with the other seven already predefined weights. Table \ref{tab:treinamento-rede-neural-caminhada-grafo-completo-pesos-solucao} shows the training result after the network was initialized. The table contains ten sets of two weights found by the oracle. The amount of times required for the backpropagation algorithm to converge after initialization is only one, that is, the network is initialized with the solution.

\begin{table}
\renewcommand{\arraystretch}{1.3}
\setlength{\tabcolsep}{5pt}
\centering
\footnotesize
\caption{Training by initialization of synaptic weights from solutions.}
\label{tab:treinamento-rede-neural-caminhada-grafo-completo-pesos-solucao}
\begin{tabular}{cccc|cccc}
\hline
\multicolumn{1}{c}{Experiment} \rule{0mm}{3mm} & \textrm{Weight1} & \textrm{Weight2} & Epoch & Experiment & \textrm{Weight1} & \textrm{Weight2} & Epoch \\ \hline
1 & -2.15 & 1.6 & 1 & 6 & -2.0 & 1.6 & 1 \\
2 & -1.75 & 1.55 & 1 & 7 & -1.85 & 1.5 & 1 \\
3 & -1.6 & 1.45 & 1 & 8 & -1.95 & 1.55 & 1 \\
4 & -1.85 & 1.55 & 1 & 9 & 1.90 & 1.55 & 1 \\
5 & -2.0 & 1.55 & 1 & 10 & -1.8 & 1.5 & 1 \\ \hline
\end{tabular}
\end{table}

Tables \ref{tab:Resultados-das-medicoes-para-janela-512-512}, \ref{tab:Resultados-das-medicoes-para-janela-1024-1024} and \ref{tab:Resultados-das-medicoes-para-janela-2048-2048} show the results of the measurements performed on the system after the walk. Once the first measurement is made, as seen previously, we obtain the state whose set of positions are solutions. So, when performing the second measurement, whatever the result it will be a solution.

\begin{table}
\renewcommand{\arraystretch}{1.3}
\setlength{\tabcolsep}{15pt}
\centering
\caption{Results of measurements for window $512\textrm{x}512$.}
\label{tab:Resultados-das-medicoes-para-janela-512-512}
\begin{tabular}{cccc}
\hline
\rule{0mm}{3mm} States & 52 & 5202 & 19591 \\ \hline
$\ket{aa}$ & 99.96\% & 99.96\% & 93.39\% \\
$\ket{ab}$ & 0.03\% & 0.03\% & 3.21\% \\
$\ket{ba}$ & 0.01\% & 0.01\% & 3.32\% \\
$\ket{bb}$ & 0.00\% & 0.00\% & 0.08\% \\ \hline
\end{tabular}
\end{table}

Table \ref{tab:Resultados-das-medicoes-para-janela-512-512} corresponds to the results of measurements for walking in windows of dimensions $512\textrm{x}512$. When $\Delta p = 0.05$ we have $52$ solutions, and with $\Delta p = 0.005$ we have $5202$ solutions and with $\Delta p = 0.0005$ we have $19591$ solutions.

\begin{figure}
    \centering
    \includegraphics[width=10cm]{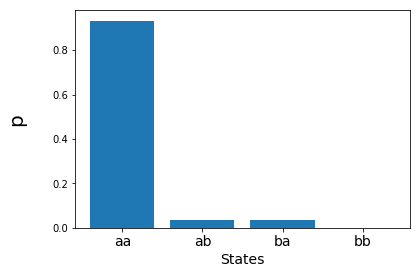}
    \caption{Distribution of probability of a walk in complete graph with 19591 solutions.}
    \label{fig:distribuicao-de-probabilidade-caminhada-grafo-completo}
\end{figure}

In Figure \ref{fig:distribuicao-de-probabilidade-caminhada-grafo-completo} it is possible to see the probability distribution of the quantum walk in complete graph referring to the fourth column of Table \ref{tab:Resultados-das-medicoes-para-janela-512-512}, where we have $19591$ solutions, and in $96.60\%$ of the measurements is successful. The result of the measures follows the distribution of quantum walk probabilities in a complete graph.

\begin{table}
\renewcommand{\arraystretch}{1.3}
\setlength{\tabcolsep}{15pt}
\centering
\caption{Results of measurements for window $1024\textrm{x}1024$}
\label{tab:Resultados-das-medicoes-para-janela-1024-1024}
\begin{tabular}{cccc}
\hline
\rule{0mm}{3mm} States & 52 & 5202 & 118649 \\ \hline
$\ket{aa}$ & 100\% & 98.95\% & 71.46\% \\
$\ket{ab}$ & 0.00\% & 0.85\% & 26.36\% \\
$\ket{ba}$ & 0.00\% & 0.20\% & 1.52\% \\
$\ket{bb}$ & 0.00\% & 0.00\% & 0.66\% \\ \hline
\end{tabular}
\end{table}

Table \ref{tab:Resultados-das-medicoes-para-janela-1024-1024} corresponds to the results of measurements for walking in windows of dimensions $1024\textrm{x}1024$. When $\Delta p = 0.05$ we have $52$ solutions, and with $\Delta p = 0.005$ we have $5202$ solutions, and with $\Delta p = 0.0005$ we have $118649$ solutions.

\begin{table}
\renewcommand{\arraystretch}{1.3}
\setlength{\tabcolsep}{15pt}
\centering
\caption{Results of measurements for window $2048\textrm{x}2048$.}
\label{tab:Resultados-das-medicoes-para-janela-2048-2048}
\begin{tabular}{cccc}
\hline
\rule{0mm}{3mm} States & 52 & 5202 & 485095 \\ \hline
$\ket{aa}$ & 99.99\% & 99.63\% & 73.28\% \\
$\ket{ab}$ & 0.01\% & 0.35\% & 24.21\% \\
$\ket{ba}$ & 0.00\% & 0.02\% & 1.83\% \\
$\ket{bb}$ & 0.00\% & 0.00\% & 0.68\% \\ \hline
\end{tabular}
\end{table}

Table \ref{tab:Resultados-das-medicoes-para-janela-2048-2048} corresponds to the results of measurements for walking in windows of dimensions $2048\textrm{x}2048$. When $\Delta p = 0.05$ we have $52$ solutions, and with $\Delta p = 0.005$ we have $5202$ solutions, and with $\Delta p = 0.0005$ we have $485095$ solutions. In all cases the probability of obtaining a solution is above $99\%$.

Another observation that must be made takes into consideration the time that procedure needs to find the solutions. In the backpropagation algorithm or same descendent gradient algorithm, the network convergence depends on several factors, such as random initialization of weights, activation functions or surface complexity, etc.  Then, \textit{a priori} it is not possible to determine the number of times the training process (or iterations) before the network can converge to a solution \citep{biamonte2017quantum}. In the case of the procedure proposed, the Equation~\ref{eq:qtd-passos-probabilidade-maxima} determines the number of iterations necessary for a set of solutions to be obtained.
\section{Conclusion}
\label{sec:conclusion}

When comparing the training of artificial neural networks using the backpropagation algorithm with the procedure developed in this research, a significant difference must be observed. The ANN training using the backpropagation algorithm searches for a minimum of an error function. In this procedure, the ANN can stagnant in a local minimum. Conversely, the process developed here is capable of finding solutions in a previously known number of iterations. 

Since the Oracle knows which positions of the walker can generate weights capable of training the neural network, the simple execution of the  quantum walk search will find the positions in the walker space that are a solution for ANN training.

As we have seen before in this work, we have studied the training of an artificial neural network of the multilayer perceptron type with two layers, three neurons, and nine weights. For reasons of limitation of processing and memory, it was necessary to apply to this procedure the same approach as initializing the training of an extreme learning machine. Therefore, the application of this procedure to find all weights for this and other functions may be proposed for future work.
\section*{Acknowledgments}
\label{sec:acknowledgments}
Acknowledgments to the Science and Technology Support Foundation of Pernambuco (FACEPE) Brazil, Brazilian National Council for Scientific and Technological Development (CNPq), and Coordena\c{c}\~{a}o de Aperfei\c{c}oamento de Pessoal de N\'{i}vel Superior - Brasil (CAPES) - Finance Code 001 by financial support for the development of this research.

\bibliographystyle{unsrtnat}






\end{document}